\documentclass{article}
\usepackage[T1]{fontenc}
\usepackage[utf8]{inputenc}
\usepackage{arxiv} 
\usepackage{amsmath,cite,url}
\usepackage{titlesec}
\usepackage{algorithm}
\usepackage{algpseudocode}
\usepackage{booktabs}
\usepackage{multirow}
\usepackage{graphicx}
\usepackage{color}

\title{ Text2midi-InferAlign:  Improving Symbolic Music Generation with Inference-Time Alignment}

\oneauthor
    { Abhinaba Roy, Geeta Puri, Dorien Herremans}
    {Singapore University of Technology and Design \\ 
    abhinaba\_roy@sutd.edu.sg, geeta\_puri@mymail.sutd.edu.sg, dorien\_herremans@sutd.edu.sg}

\def\authorname{F. Author, S. Author, and T. Author}

\newcommand{\model}{\textbf{\texttt{Text2midi-InferAlign}}}

\makeatletter
\def\@maketitle{%
  \newpage
  \null
  \vskip 2em%
  \begin{center}%
    {\Large \bfseries \@title \par}%
    \vskip 1.5em%
    {\normalsize \lineskip 0.5em%
     \begin{tabular}[t]{c}%
       \@author
     \end{tabular}\par}%
    \vskip 1em%
  \end{center}%
  \par
  \vskip 1.5em}
\makeatother

\begin{document}

\maketitle

\begin{abstract}
We present \model, a novel technique for improving symbolic music generation at inference time. Our method leverages text-to-audio alignment and music-structural alignment rewards during inference to encourage the generated music to be consistent with the input caption. Specifically, we introduce two objectives scores: a \textit{text-audio consistency score} that measures rhythmic alignment between the generated music and the original text caption, and a \textit{harmonic-consistency score} that penalizes generated music containing notes inconsistent with the key. By optimizing these alignment-based objectives during the generation process, our model produces symbolic music that is more closely tied to the input captions, thereby improving the overall quality and coherence of the generated compositions. Our approach can extend any existing autoregressive model without requiring further training or fine-tuning. We evaluate our work on top of Text2midi - an existing text-to-midi generation model, demonstrating significant improvements in both objective and subjective evaluation metrics.

\end{abstract}

\section{Introduction}\label{sec:introduction}
The field of multimodal generative models has witnessed remarkable progress, particularly in tasks involving text and images \cite{wang2024comprehensive}. However, generation of symbolic music from textual descriptions remains a relatively unexplored challenge, requiring models to understand the intricate relationships between language and music. Recently, the Text2midi model \cite{bhandari2024text2midi} was introduced, representing the first end-to-end approach for generating midi files from text captions using large language models. Such models leverage the availability of textual data and the success of large language models (LLMs), using a pretrained LLM encoder to process captions and an autoregressive transformer decoder to produce midi sequences that reflect the provided descriptions \cite{bhandari2024text2midi}. 

\begin{figure}[t]
\includegraphics[width=8cm]{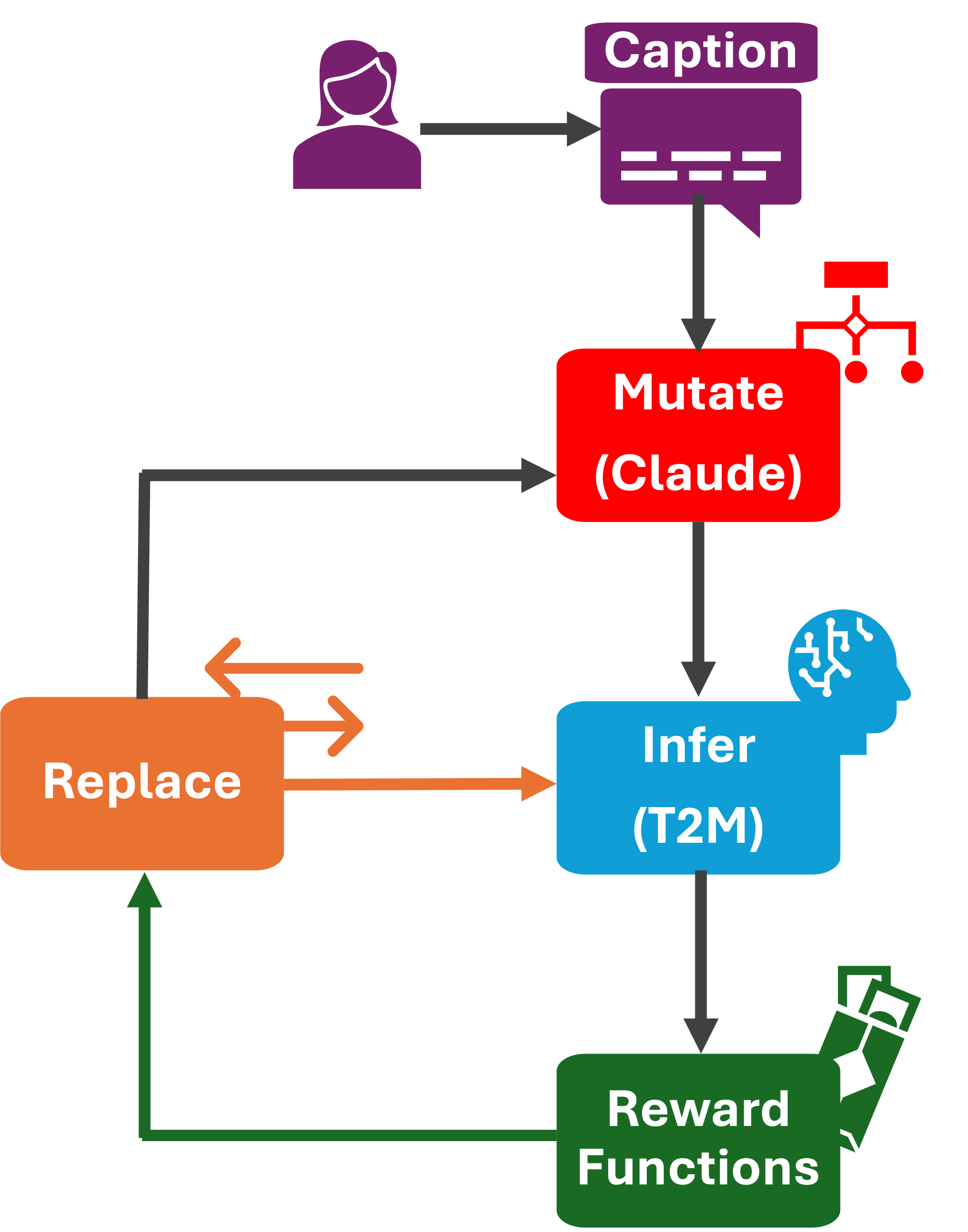}
\caption{Overview of \model, using Text2midi for inference and reward score(s) for alignment. }
\label{fig:overview}
\end{figure}

While autoregressive models are adept at reducing the cross-entropy between music and text embeddings in the decoder, this approach does not always guarantee adherence to music-structural integrity \cite{ji2020comprehensive}. Musicality, which determines if a piece of music sounds  "nice" (i.e., expressive and aesthetically pleasing) to the listener, is not explicitly optimized for or even measured in these models. Additionally, during generation, it is crucial that the symbolic music aligns with the input text caption, yet current models often lack explicit mechanisms to ensure this alignment during inference.

The concept of "niceness" is an important consideration in music generation and could be measured through user preference. In the field of large language models (LLM), this is often addressed through reinforcement learning based on human feedback, where evaluators directly judge the quality and relevance of the generated content and provide rewards accordingly \cite{sun2024llm}. These reward scores are then used to fine-tune the language model, making its responses more human-like. Although LLMs have explored related ideas, only \cite{cideron2024musicrl} have applied this human preference feedback to fine-tune text-to-music generation models. However, the reliance on human input imposes significant cost and limits the model's ability to adapt to diverse user preferences. This also requires expensive re-training of the original model.

Recent studies have demonstrated the effectiveness of inference-time alignment techniques in improving the performance of large language models on general-purpose response generation tasks \cite{hung2024inference} as well as reasoning \cite{Achiam2023GPT4TR}. These approaches circumvent the need for retraining the underlying model, instead generating more human-value aligned responses on the fly. Motivated by these advancements, we introduce \model \footnote{code and examples are available at https://github.com/AMAAI-Lab/t2m-inferalign} a novel method that aims to enhance symbolic music generation through the application of inference-time alignment.  In our work, we employ a similar inference-time alignment approach as \cite{hung2024inference}, framing the problem as a reward-guided tree search. Tree search algorithms alternate between two key stages: exploration and exploitation \cite{browne2012survey}. At any given state, the exploration phase determines which potential options to investigate further, while the exploitation phase selects the most promising options based on a pre-defined objective and corresponding reward metric. Motivated by this approach, we employ a two-pronged methodology to enhance the tree search process. Firstly, we incorporate an "instruction mutation" technique that leverages language models to generate variations of the original caption. This promotes exploratory behavior by expanding the search space to encompass potentially more relevant textual descriptions. Secondly, we utilize \textit{reward-guided beam replacement} \cite{hung2024inference} to encourage exploitative behavior, wherein we select the most promising options based on pre-defined objective functions and corresponding reward metrics. Here, we introduce two distinct objectives: a \textit{text-audio consistency score} that measures alignment between the generated music and the original text caption, and a \textit{harmonic-consistency score} that penalizes generated music containing notes inconsistent with the key. By optimizing these alignment-based rewards during the generation process, our approach aims to improve the overall quality and coherence of the symbolic music produced by Text2midi \cite{bhandari2024text2midi}.

In summary, the key contributions of our work are:
\begin{enumerate}
    \item We propose \model, a novel inference-time alignment method that enhances the quality of symbolic music generation.
    \item We incorporate \textit{caption mutation} to promote exploration, generating variations of the original caption to expand the search space. 
    \item We introduce two objectives scores - \textit{text-audio consistency} and \textit{harmonic-consistency} to encourage exploitation, selecting the most promising options.
    \item We demonstrate that \model\ can significantly improve both objective and subjective evaluation metrics for symbolic music generation. 
\end{enumerate}

\section{Related Work}
Our work builds upon research in several key areas: automatic music generation, text-to-music generation, and inference-time alignment. Music generation has a rich history, with early approaches relying on rule-based systems and Markov models \cite{hiller1957musical, farbood2001analysis, funk2018musical} to generate musical sequences. In recent times, with the advent of deep learning, automatic music composition has seen great interest. Researchers have found that models based on recurrent neural networks, such as long short-term memory models (LSTM), can effectively capture long-term structure in the generated music sequences \cite{zixun2021hierarchical}. Prominent examples of LSTM-based models include BachBot \cite{liang2016bachbot} and DeepBach~\cite{hadjeres2017deepbach}, which demonstrate proficiency in generating music in the style of Bach. Furthermore, the introduction of the Transformer architecture, coupled with the availability of expanding midi datasets, has enabled the development of more powerful and expressive models, such as Music Transformer \cite{huang2018musictransformer} and Museformer \cite{yu2022museformer}. The Museformer model, in particular, leverages the fine-grained and coarse-grained attention mechanisms of Transformers to produce high-quality and relatively lengthy music sequences. 

Recently, a number of of works have proposed approaches for music generation from text \cite{melechovsky2023mustango, copet2023simple, huang2023noise2music}. These models consider a free-form text prompt and  leverage pretrained large language text encoders, such as FLAN-T5~\cite{chung2024scaling}, and feed the resulting embedded text representation as input to an audio decoder. Some models generate shorter fragments using diffusion \cite{melechovsky2023mustango}, whereas others allow longer fragments through an autoregressive transformer approach \cite{copet2023simple}. Generative approaches for \emph{symbolic} music generation also has seen recent developments in recent times \cite{pasquier2025midigptcontrollablegenerativemodel, bhandari2024text2midi, lu2023musecocogeneratingsymbolicmusic}. MuseCoco \cite{lu2023musecocogeneratingsymbolicmusic} generates symbolic music from text descriptions using a two stage framework: 1) text-to-attribute understanding and 2) attribute-to-music generation - provides more control over music elements. Moreover, chatGPT4o\footnote{https://chatgpt.com} also generates midi files based on input text promt. However, quality of generation is rather questionable as the generated midi files barely contain notes. This is also indicated in \cite{lu2023musecocogeneratingsymbolicmusic}. 
Another approach for developing symbolic music from text captions is from \cite{zhang2020butter}, who developed the BUTTER (Better Understanding of Text To Explanation and Recommendation) model. This model generates music in ABC format\footnote{https://abcnotation.com} based on given text instruction. Like MuseCoco, this model also takes a two-stage approach and predicts musical attributes by looking into the text caption to generate folk music. On the contrary, Text2midi \cite{bhandari2024text2midi} proposed an end-to-end transformer based autoregressive architecture that is quite fast and easy to modify. We choose Text2midi as baseline model for our approach. 

Inference time alignment has extensively been studied in the field of LLMs. Methods such as \cite{sessa2024bond, Achiam2023GPT4TR, Khanov2024ARGSAA, hung2024inference} have demonstrated effectiveness in inference time alignment in LLM response generation. However, to the best of our knowledge, no work has addressed inference time alignment for symbolic music generation. In this work, we follow a similar approach as \cite{hung2024inference}, who pose alignment as a reward guided tree search. We formulate reward functions based on two objectives and seek to \textit{explore} and \textit{exploit} the search space to come up with best possible set of midi tokens generated by Text2midi that aligns most with our objectives. 

\section{Method}
This section offers a detailed description of \model, our proposed inference-time alignment technique for improving symbolic music generation. We first present the mathematical formulation of the problem and then outline the overall algorithm.
Given a text caption $x$ and a pre-trained Text2midi model $f_{\theta}$ parameterized by $\theta$, our goal is to generate a midi sequence $y$ that aligns with the caption $x$ while adhering to music-structural integrity. 
Let $f_{\theta}$ be the Text2midi model that takes a text caption $x$ as input and generates a midi sequence $y$.
\begin{equation}
    y = f_{\theta}(x)
\end{equation}

where $y = [y_1, y_2, ..., y_\mathcal{N} ]$ is the generated midi sequence of length $\mathcal{N}$.
Since $f_{\theta}$ is an autoregressive model, the probability of generating the midi sequence $y$ can be expressed as:
\begin{equation}
    p(y|x; \theta) = \prod_{t=1}^{\mathcal{N}} p(y_t | y_{1:t-1}, x; \theta)
\end{equation}
We treat midi token generation through \model\  as a tree search problem to find the optimal set of states $[1,2,3,...,\mathcal{S}]$ that generate the $\mathcal{N}$ midi tokens. A state $s$ ($1 \leq s\leq S$) is a set of generated midi tokens $s = [y_1,y_2, ..., y_n]; n \leq \mathcal{N}$. Given an instruction $x$, we aim to search for $s* = [y_1*, y_2*, ..., y_t*]$ that maximizes a reward score $\mathcal{R}(s*, x)$. If states $s_1$ and $s_2$ are reached following instructions $x_1$ and $x_2$, respectively, and $\mathcal{R}(s_1, x)>\mathcal{R}(s_2, x_2)$, then $x_1 \succ x_2$. We employ \textit{exploitation} and \textit{exploration} in our tree search. Reward exploration mutates a given caption into several captions using a language model, guiding independent search processes. This encourages diverse exploration of the state-space, potentially discovering higher reward states. After this, we focus on leveraging
high-reward states discovered during the search process and continue the search process from there.  

\subsection{Objectives and Reward Functions}

We define two objectives to ensure high musical quality as well as alignment with the input text caption. Each of these objectives employs a reward function to evaluate and score a particular state. The two objectives and the corresponding reward functions are as follows -

(a) \textbf{Text-Audio Consistency}: To ensure that the generated midi sequence is rhythmically and semantically aligned with the text caption $x$. We use clap score \cite{laionclap2023} to measure the similarity between the text caption and the generated midi tokens. Mathematically, 
\begin{equation}
    \mathcal{R}_{a}(s, x) = \texttt{CLAP}([y_1,y_2, ..., y_n], x)
\end{equation}
Here, $n \leq \mathcal{N}$.

 (b) \textbf{Harmonic Consistency}: This objective is to ensure the generated music adheres to basic music theory principles, by penalizing notes in the generated sequence inconsistent with the specified key \cite{guo2019midi, chew2000towards}.

\begin{equation}
    \mathcal{R}_{h}(s,x) = 1 - \frac{\texttt{\#off key note}}{\texttt{\#notes}}
    \label{eq:off}
\end{equation}

We define the composite reward function as the weighted sum of the two rewards:
\begin{equation}
    \mathcal{R}(s,x) = \alpha \mathcal{R}_{a} + \beta \mathcal{R}_{h}
    \label{eq:r}
\end{equation}

where $\alpha$ and $\beta$ are hyperparameters to balance the relative importance of the two rewards.

\subsection{Exploration}
During \textit{exploration}, we use caption mutation to generate captions $\{x_1, x_2, ..., x_\mathcal{T}\}$ to generate $\mathcal{T}$ instructions to guide the search process. This results in reaching $\{s_1, s_2, . . . , s_\mathcal{T} \}$ where $s_i$ represents the state reached with $x_i$. The highest valued state based on the reward function $\mathcal{R}$ is the most favorable state. 
$s*=\text{argmax}_{s\in {s_1,s_2,...,s_\mathcal{T}}} \mathcal{R}(s, x)$. For all reward function calculations, we calculate with respect to the original caption $x$, in order to stop mutations deviate too much from the original caption. 

We use Claude-3-Haiku\footnote{https://www.anthropic.com/claude/haiku} LLM to generate the caption mutations. Specifically, we use in-context-learning to prompt the LLM with an example of mutations from a captions and ask it to generate $\mathcal{T}$ variations of an input caption. 

\subsection{Exploitation}
During the \textit{exploitation} stage, we use reward function scores to replace the low-scoring states with potentially high-scoring states. Formally, if we have a tuple of states $s_1, s_2, , ..., s_q$ ordered by function scores such that $\mathcal{R}(s_1,x)> \mathcal{R}(s_2,x)> ... > \mathcal{R}(s_q,x)$ the replacement can be defined as 
\begin{equation}
    \Phi(s_1, s_2, , ..., s_q) = (s_1, s_2, , ..., s_k, r_1, ..., r_{q-k})
\end{equation}
Where $k<q$ and $r_i \in {s_1, . . . s_k}$, for all $i = 1,..., q-k$. We randomly replace any state that is not among the top $k$ highest scoring states. In practice, We apply this replacement operation to every $m$ tokens generated until the maximum number of tokens are generated, indicating the end of the search process. During replacement we also preserve the instructions that result in high scoring reward functions. Algorithm ~\ref{al:1} describes our approach.


\begin{algorithm}
    \caption{\model}
    \begin{algorithmic}[1]
        \Require $x$: Original caption, $\mathcal{R}(\cdot)$: Reward function, 
        $m$: Replacement period, $Z$: \# of mutation cycles, 
        $\mathcal{T}$: \# of mutations, $\tau$: \# of replacement cycles per mutation
        \State $A \gets \{x\}$ \Comment{Initialize candidate set}
        \For{$i = 1, 2, \dots, Z$} \Comment{Mutation cycles}
            \State $x_{\text{candidate}} \sim A$ \Comment{Select candidate}
            \State $S^{(1)} \gets \{ s_j \mid s_j = \phi, j \in \{1, \dots, \mathcal{T}\} \}$
            \State $x_{\text{mutated}} \gets \text{Mutate}(x_{\text{candidate}}, \mathcal{T})$
            \For{$t = 2, 3, \dots, \tau$} \Comment{Replacement cycles}
                \State $S^{(t)} \gets \text{Infer}(S^{(t-1)}, x_{\text{mutated}}, m)$
                \State $S^{(t)} \gets \text{Replace}(S^{(t)}, \mathcal{R}(\cdot))$
                \State $S_{\text{top-k}}^{(t)} \gets \arg\max_{-k, s \in S^{(t)}} \mathcal{R}(s, x)$
                \State $x_{\text{top-k}}^{(t)} \gets \{ x_{\text{mutated}}[s] \mid s \in S_{\text{top-k}}^{(t)} \}$
            \EndFor
            \State $x_{\text{top-k}} \gets \text{Most frequent elements in } x_{\text{top-k}}^{(2:\tau)}$
            \State $A \gets \{ x_i \mid x_i \succ x_{\text{candidate}}, x_i \in x_{\text{top-k}} \}$
        \EndFor
    \end{algorithmic}
    \label{al:1}
\end{algorithm}
\section{Experimental Setup}
\label{sec:ex}
In our study, we investigate the effectiveness of inference-time alignment in enhancing symbolic music generation, focusing on both objective metrics and subjective evaluations. In this section, we discuss the experimental setup. The results are reported in the next section. 

\subsection{Baseline \& Test set}
We use a pre-trained text2midi model \cite{bhandari2024text2midi} to demonstrate the effectiveness of our inference-time alignment technique for improving symbolic music generation. The model was trained on the MidiCaps dataset \cite{melechovsky2024midicaps}, which comprises pairs of text captions and midi files. Accordingly, for both subjective and objective evaluations, we employ examples from the MidiCaps test set. For the subjective assessment, we utilize 8 test examples. For objective evaluation, we consider 50\% of the MidiCaps test set.
\subsection{Objective Evaluation Metrics}
\label{sec:obj}
For objective evaluation, we follow similar metrics as those used in \cite{bhandari2024text2midi}. These measure how much long-term structure and patterns the music contains as well as how similar the input text and generated music are and how close some of the important musical features are to those of the ground truth. The metrics are as follows:
\paragraph{Compression ratio} uses the COSIATEC algorithm \cite{meredith2013cosiatec} as used by \cite{chuan2018modeling} to quantify the recurring patterns and long-term structure and  within the music in midi files.

\paragraph{CLAP} or Contrastive Language-Audio Pretraining \cite{laionclap2023} captures a joint latent representations of audio and text samples. We use an improved CLAP checkpoint (LAION CLAP\footnote{https://huggingface.co/laion/larger\_clap\_music\_and\_speech}), specifically trained on music and speech. We extract latent representations of an input caption and the synthesized audio of the midi file. We then use the cosine similarity to measure of similarity between caption and audio (synthesized from midi).

\paragraph{Feature wise comparison} calculates four features extracted from generated midi files and their related ground truth as used by \cite{melechovsky2023mustango}:
\begin{itemize}
\item Tempo Bin (TB): represents the proportion of extracted tempo that falls within the predetermined tempo bins derived from the ground truth midi files. Tempo bin boundaries are defined at 40, 60, 70, 90, 110, 140, 160, and 210 beats per minute. The tempo is extracted with music21 \cite{cuthbert2010music21}. 
%
\item Tempo Bin with Tolerance (TBT): measures the proportion of generated midi files where the predicted tempo falls within the ground truth tempo bin or an adjacent bin, relative to the total number of files.
\item Correct Key (CK): the ratio of midi files of which the extracted key (using music21 \cite{cuthbert2010music21}) matches the extracted of the ground truth key midi file.

\item Correct Key with Duplicates (CKD): exhibits a slightly more lenient approach compared to the previous metric. It quantifies the proportion of generated midi files where the extracted key matches either the key of the ground truth midi file or an equivalent key.
\end{itemize}
\subsection{Listening Test}
\label{sec:lt}
For subjective evaluation of our model, we conducted a listening study utilizing PsyToolkit \cite{stoet2010psytoolkit}. We chose five captions from the test set of MidiCaps and three free-text captions to generate two sets of midi files by Text2midi and \model. Participants were asked to choose the preferred music (synthesized from midi) based on two criteria: \textbf{musical quality} and \textbf{text-audio matching}. Captions from the MidiCaps test set contain more musical feature details whereas free-text captions are shorter and contain less musical informations.

A MidiCaps test-set example: \textit{"A melodic electronic song with ambient elements, featuring piano, acoustic guitar, alto saxophone, string ensemble, and electric bass. Set in G minor with a 4/4 time signature, it moves at a lively Presto tempo. The composition evokes a blend of relaxation and darkness, with hints of happiness and a meditative quality." }. 

An example of free-text caption: \textit{"Upbeat acoustic guitar tune with a warm and cheerful feel."}

All captions-generated pair are available at our GitHub page\footnote{https://github.com/AMAAI-Lab/t2m-inferalign}.
\subsection{Implementatinal Details}
We build our approach on top of Text2midi. We use python and modify the code from their code\footnote{https://github.com/AMAAI-Lab/Text2midi}. We use single Nvidia L40S GPU to run all inferences reported in this work. Using GPU is advantageous for faster parallelization especially during the \textit{exploration} phase. For hyperparameter optimization, we setup two ablation studies to determine optimal values of replacement period $m$ and the number
of mutations $\mathcal{T}$ (Eq. ~\ref{al:1}). We fix $\alpha = 1$ and $\beta = 5$ in Eq. ~\ref{eq:r}. We will publish our code upon acceptance. 

\section{Results}
We discuss the results of our experiments here. For all evaluation tasks, we generate midi with 2,000 tokens.
\subsection{Objective Evaluation}
We note in algorithm \ref{al:1}, our approach is dependent on two hyperparameters: replacement period $m$ and the number of mutations $\mathcal{T}$. We setup an ablation study to find out the effect of  these two hyperparameters.  For the ablation studies, we consider 16 randomly selected examples from MidiCaps test set.

\textbf{Effect of $m$}: We calculate TB, TBT, CK, CKD for midi generations from 16 randomly selected captions from MidiCaps test set using \model. We consider four values of $m$ -  \texttt{100}, \texttt{500}, \texttt{1000}, and \texttt{2000}. Please note, we generate a maximum of 2000 tokens, hence choosing $m$=\texttt{2000} is equivalent to Best-of-N approach. For all generation scenarios here, we fix $\mathcal{T}$ to 3. Results are shown in Table \ref{tab:1}. Numbers indicate average 16 chosen captions. We can notice $m$=\texttt{100} performs best compared to the rest. Although Best-of-N ($m$=\texttt{2000}) performs equally well for TB,TBT and CK; $m$=\texttt{100} shows better CKD (43.75 vs 37.5). For all consecutive experiments, we consider $m$=\texttt{100}.

\begin{table}[h!]
    \centering

    \begin{tabular}{c|cccc}
    \toprule
      $m$ &  \texttt{100} & \texttt{500} & \texttt{1000} & \texttt{2000} \\
      \midrule
    TB (\%) $\uparrow$ & 37.50 & 37.50 & 31.25 & 37.50\\
     TBT (\%) $\uparrow$    & 62.50 & 62.50 & 56.25 & 62.50 \\
     CK (\%) $\uparrow$ & 37.50 & 31.25  & 43.75 & 37.50\\
     CKD (\%) $\uparrow$ & 43.75 & 31.25 & 43.75 & 37.50 \\
     \bottomrule
    \end{tabular}%

    \caption{Effects of replacement period $m$. Numbers are average results for 16 randomly chosen captions from the MidiCaps test set. TB: Tempo Bin; TBT: Tempo Bin with Tolerance; CK: Correct Key; CKD: Correct Key with Duplicates; $\uparrow$: higher score better.}
    \label{tab:1}
\end{table}

\textbf{Effect of $\mathcal{T}$}: Similar to the previous ablation study, we calculate  TB, TBT, CK, CKD for 16 midi generations (the same captions as used in the previous ablation study) using \model. We consider $\mathcal{T}$ = \texttt{1}, \texttt{3}, and \texttt{5}. $\mathcal{T}$ = \texttt{1} is the edge case where practically no mutation is happening. The results are shown in Table \ref{tab:2}. Although TB, TBT  fares equally well across all values of $\mathcal{T}$; CK and CKD increases gradually, with best overall performance being achieved by $\mathcal{T}=\texttt{5}$. Choosing a higher number for $\mathcal{T}$ becomes computationally expensive, hence we stick with $\mathcal{T} =\texttt{5}$ and do not investigate further.  

\begin{table}[h!]
    \centering

    \begin{tabular}{c|ccc}
    \toprule
      $\mathcal{T}$ &  \texttt{1} & \texttt{3} & \texttt{5} \\
      \midrule
    TB (\%) $\uparrow$ & 37.50 & 37.50 & 37.50 \\
     TBT (\%) $\uparrow$    & 62.50 & 62.50 & 62.50  \\
     CK (\%) $\uparrow$ & 37.50 & 37.50  & \textbf{50.00} \\
     CKD (\%) $\uparrow$ & 37.50 & 43.75 & \textbf{50.00} \\
     \bottomrule
    \end{tabular}%

    \caption{Effects of number of mutations $\mathcal{T}$. Numbers are average results for 16 randomly chosen captions from the MidiCaps test set. TB: Tempo Bin; TBT: Tempo Bin with Tolerance; CK: Correct Key; CKD: Correct Key with Duplicates; $\uparrow$: higher score better.}
    \label{tab:2}
\end{table}

\textbf{Comparison with Text2midi}: To properly establish the efficacy of our approach, we compare with Tex2midi on all six metrics described in Sec. \ref{sec:obj}. Based on the ablation studies done before, we fix $m=100$ and $\mathcal{T}=\texttt{5}$. The results are shown in Table~\ref{tab:3}. We improve on all six metrics. The CLAP score sees a considerable 29.4\% improvement (from 0.17 to 0.22). TB value increases almost 32.5\% (from 29.73 to 39.41). Interestingly, key matching (CK) also improved from 37.5 to 50.0. This indicates that text-audio and harmonic alignment affects other areas of musicality, such as long-term structure (CR), as well. 
\begin{table}[h!]
    \centering

    \begin{tabular}{c|cc}
    \toprule
     &  Text2Midi & \model \\
      \midrule
    CR $\uparrow$ & 2.16 & \textbf{2.31} \\
    CLAP $\uparrow$ &  0.17 & \textbf{0.22}  \\
    \midrule
    TB (\%) $\uparrow$ & 29.73 & \textbf{39.41}  \\
     TBT (\%) $\uparrow$    & 60.06 & \textbf{62.59} \\
     CK (\%) $\uparrow$ & 13.59 & \textbf{29.80}   \\
     CKD (\%) $\uparrow$ & 16.66 & \textbf{32.54}  \\
     \bottomrule
    \end{tabular}%

    \caption{Comparison between Text2midi and \model . Numbers are average results for 50\% of the captions from the MidiCaps test set. CR: Compression ratio; CLAP: CLAP score; TB: Tempo Bin; TBT: Tempo Bin with Tolerance; CK: Correct Key; CKD: Correct Key with Duplicates; $\uparrow$: higher score better.}
    \label{tab:3}
\end{table}

\begin{table*}[h]
    \centering
    \renewcommand{\arraystretch}{1.2}
    \begin{tabular}{lcc}
        \toprule
        & Text2midi (\%) & \model (\%) \\
        \midrule
         Music quality & 31.25 & \textbf{68.75} \\
        Text-audio match & 41.67 & \textbf{58.33} \\
        \bottomrule
    \end{tabular}
    \caption{Comparison of Text2midi and \model\ in terms of music quality and text matching.}
    \label{tab:comp1}
\end{table*}

\begin{table*}[h]
    \centering
    \renewcommand{\arraystretch}{1.2}
    \begin{tabular}{lcc}
        \toprule
        & Text2midi (\%) & \model (\%) \\
        \midrule
         MidiCaps caption & 48.33 & \textbf{51.67} \\
        Free text caption & 27.78 & \textbf{72.22} \\
        \bottomrule
    \end{tabular}
    \caption{Comparison of Text2midi and \model\ in terms of type of captions.}
    \label{tab:comp2}
\end{table*}

\subsection{Listening Test}
A total of 11 listeners were involved in the subjective listening test. The results are shown in Table ~\ref{tab:comp1} and ~\ref{tab:comp2},  which display the percentage of responses preferring each option. Table ~\ref{tab:comp1} showcases our performance based on music quality and text-audio matching.  Notably, our approach is significantly preferred for its music quality (68.75\% vs 31.25\%), lending support to the hypothesis that our alignment approach enhances the overall musical quality in a comprehensive manner.

Table ~\ref{tab:comp2} shows the breakdown of performance by type of captions (Sec. ~\ref{sec:lt}). We notice free-text captions favour hugely our model.  This is due to the ability of our approach to mutate the original captions and explore a wider variety of possibilities, which in turn encourages and supports the Text2midi to generate higher quality midi files. Our flexible alignment method enables the model to better accommodate the open-ended nature of free-text descriptions, leading to more coherent and musically compelling outputs. On the other hand, MidiCaps captions contain more detailed musical information, which inhibits extensive exploration and mutation during the inference time. However, our alignment approach can still be beneficial in such cases. 

From these results, we can see that our method can also help users find better quality or more suitable musical outputs, even when they are not entirely certain about the specific musical qualities they are looking for, which is the case with free-flow captions. The ability to adapt to different types of textual descriptions, from open-ended free-text to more constrained MidiCaps captions, demonstrates the versatility of our alignment technique and its potential to enhance the overall text-to-midi generation experience.
\section{Conclusion}
In this study, we introduced \model, an inference-time alignment method for text-to-midi generation, addressing key challenges in ensuring text-audio consistency, as well as harmonic consistency in the generated symbolic music. Our approach is inspired by tree search-based alignment techniques, which we adapt to the unique characteristics of symbolic music generation. Through empirical evaluation, we demonstrate that our method enhances semantic coherence and musical quality without necessitating additional training, rendering it a lightweight yet effective enhancement.
Prospective research could explore integrating reinforcement learning to enable more fine-grained alignment, extending our approach to more musical elements. We believe our contributions pave the way for more expressive AI-driven music composition.

\section{Acknowledgment}
This work has received support from SUTD’s Kickstart Initiative under grant number SKI 2021 04 06 and MOE under grant number MOE-T2EP20124-0014. We acknowledge the use of ChatGPT for grammar refinement and paraphrasing.

\bibliography{ISMIRtemplate}

\end{document}